\documentclass[12pt]{article}
\usepackage{epsf}
\usepackage{aas_macros}
\usepackage{amssymb}
\usepackage{amsmath}
\usepackage{braket}
\usepackage{mathrsfs}
\usepackage{mathtools}
\usepackage{array}
\usepackage{fancyhdr}
\usepackage[dvipdfmx]{graphicx}
\usepackage{color}
\usepackage{cite}
\usepackage{bm}
\usepackage{url}
\usepackage{fancyhdr}
\usepackage{comment}
\usepackage[colorlinks,citecolor=blue]{hyperref}
\renewcommand{\thefootnote}{\#\arabic{footnote}}

\renewcommand{\thefootnote}{\fnsymbol{footnote}}
\def\thefootnote{\fnsymbol{footnote}}

\makeatletter

\@addtoreset{equation}{section}
\makeatother

\newcommand{\mrm}{\mathrm}
\newcommand{\nn}{\nonumber}

\newcommand{\ga}{\gamma}

\newcommand{\si}{\sigma}

\newcommand{\ka}{\kappa}
\newcommand{\al}{\alpha}
\newcommand{\be}{\beta}
\newcommand{\de}{\delta}
\newcommand{\De}{\Delta}

\newcommand{\Om}{\Omega}

\newcommand{\cd}{\cdot}

\newcommand{\x}{\bm{x}}
\newcommand{\y}{\bm{y}}

\newcommand{\asec}[2]{#1^{''}\!\!.#2}

\newcommand{\BEA}{\begin{eqnarray}}
\newcommand{\EEA}{\end{eqnarray}}
\def\vec#1{\boldsymbol #1}
\begin{document}

\begin{titlepage}

\begin{center}

\vskip .75in

{\Large \bf Probing warm and mixed dark matter models \vspace{2mm} \\ using lensing shift power spectrum}

\vskip .75in

{
Kaiki~Taro~Inoue$\,^1$,  Takumi~Shinohara$\,^2$,  Teruaki~Suyama$\,^3$,  and  Tomo~Takahashi$\,^4$
}

\vskip 0.25in

{\em
$^{1}$Faculty of Science and Engineering, Kindai University, Higashi-Osaka, 577-8502, Japan
\\[6pt]
$^{2}$Graduate School of Science and Engineering, Saga University, Saga 840-8502, Japan
\vspace{2mm}\\
$^{3}$Department of Physics, Tokyo Institute of Technology, 2-12-1 Ookayama, Meguro-ku,Tokyo 152-8551, Japan
\vspace{2mm}\\
$^{4}$Department of Physics, Saga University, Saga 840-8502, Japan
}

\end{center}
\vskip .5in

\begin{abstract}

We argue that the lensing power spectrum of astrometric shift (lensing shift power spectrum) is a powerful tool of the clustering property of dark matter on subgalactic scales. First we give the formalism to probe the nature of dark matter by using the lensing shift power spectrum. Then, leveraging recent measurements of the lensing shift power spectrum on an angular scale of approximately 1~arcsec towards the gravitationally lensed quasar MG\,J0414+0534 at the redshift of $z_S=2.639$, we place constraints on the mass of warm dark matter (WDM) particles $m_{\rm WDM}$ and their fraction in a mixed dark matter (MDM) model $r_{\rm WDM}$, in which WDM and cold dark matter coexist. Although the constraint derived from the above single lensing system is not as strong as the existing constraints, as we show in this paper, the lensing shift power spectrum has a great potential to obtain much tighter constraints on WDM and MDM models through future observations, highlighting the importance of well-controlled systematic error considerations for achieving enhanced precision.

\end{abstract}

\end{titlepage}

\renewcommand{\thepage}{\arabic{page}}
\setcounter{page}{1}
\renewcommand{\thefootnote}{\#\arabic{footnote}}
\setcounter{footnote}{0}

%%%%%%%%%%%%%%%%%%%%%%%%%%%%%%%%%%%%%%%%%
\section{Introduction \label{sec:intro}}
%%%%%%%%%%%%%%%%%%%%%%%%%%%%%%%%%%%%%%%%%

Many years have passed since the existence of dark matter (DM) was suggested (for reviews from a historical perspective, see, e.g.,  \cite{Bertone:2016nfn,deSwart:2017heh}). Since then, vast evidence for DM has been accumulated and now we know that approximately 25 \% of the present Universe consists of DM (see, e.g., results from Planck satellite \cite{Planck:2018vyg}). Although the identity of DM is still a mystery, cold dark matter (CDM) paradigm has been very successful in explaining a wide range of observations on large scales. 

However, it has also been argued that the CDM-based simulations may raise some inconsistencies with observations of structures on small scales such as the missing satellite  problem \cite{Moore:1999nt,Klypin:1999uc}, the cusp-core problem \cite{deBlok:2001rgg,deBlok:2002vgq} and too-big-to-fail problem \cite{Boylan-Kolchin:2011qkt,Boylan-Kolchin:2011lmk}. Although these problems might be resolved at least partially once one properly takes account of baryonic effects (e.g., \cite{Wadepuhl:2010ib,Parry:2011iz,Governato:2012fa}), there have been many works trying to address these issues by invoking non-cold dark matter scenarios such as warm dark matter (WDM) in which small scale structure can be suppressed due to free-streaming effect. Indeed, WDM can naturally arise in some particle physics models such as sterile neutrinos \cite{Dodelson:1993je,Shi:1998km}, gravitino \cite{Pagels:1981ke,Bond:1982uy}, axino \cite{Covi:2009pq} and so on.  Motivated by the perspectives of these considerations, constraints on the masses of WDM $m_{\rm WDM}$ have been studied in various works by using Lyman-$\alpha$ forest (for a recent work, see \cite{Irsic:2017ixq}), strong gravitational lensing \cite{inoue-etal2015, Hsueh:2019ynk,Gilman:2019nap}, satellite galaxies of the Milky way \cite{Newton:2020cog,Dekker:2021scf} and so on, which put a lower bound on $m_{\rm WDM}$ as $m_{\rm WDM} > 2-5 \,\,{\rm keV}$ at 95 \% C.L. By combining some of these observations, tighter bounds on $m_{\rm WDM}$ have been obtained as $m_{\rm WDM} > 6.048\,\, {\rm keV} \,\, (95 \% \,\, {\rm C.L.})$ from the combination of strong gravitational lens, Lyman-$\alpha$ and Milky Way satellites \cite{Enzi:2020ieg} and $m_{\rm WDM} > 9.7\,\, {\rm keV} \,\, (95 \% \,\, {\rm C.L.})$ from strong gravitational lens and Milky Way satellites \cite{Nadler:2021dft}.

Since constraints on $m_{\rm WDM}$ have become more stringent, the ability of WDM in resolving the small-scale problems has been challenged \cite{Schneider:2013wwa}. However, a simple extension such as a mixed dark matter (MDM) scenario in which cold and warm DM co-exist may make the WDM paradigm still viable as a solution to the small-scale crisis, and the small scale structure can be suppressed in a different manner from the one in the pure WDM model. The MDM scenario can also be realized in particle physics theories (see, e.g., \cite{Ibe:2013pua,Harada:2014lma}), which stimulated works to constrain the WDM mass and the fraction of WDM in MDM using Lyman-$\alpha$ forest, gravitational lensing and so on \cite{Maccio:2012rjx,Anderhalden:2012jc,Boyarsky:2008xj,Diamanti:2017xfo,Kamada:2016vsc} (see also, e.g. \cite{Giri:2022nxq}, expected constraints from future 21cm observations).

So far, many studies have used the flux ratios of quadruply lensed quasars to constrain the properties of dark matter on subgalactic scales \cite{mao1998,metcalf2001, chiba2002, Dalal:2001fq,keeton2003, hisano2006,xu2009,inoue-etal2015} (see also \cite{Hsueh:2019ynk,Gilman:2019nap} for recent works). Actually measured flux ratios can be used to estimate the lensing convergence power spectrum \cite{Bayer:2018vhy} (see also \cite{Hezaveh:2014aoa,DiazRivero:2018oxk,Brennan:2018jhq,Cyr-Racine:2018htu,Chatterjee:2017orx}), from which non-CDM models could also be constrained.

Other studies have used astrometric shifts of lensed extended images \cite{treu-koopmans2004, koopmans2005, vegetti2009, vegetti2010, chantry2010, vegetti2014}. Anomalous astrometric shifts due to a dwarf-galaxy sized halo have been detected in the lensed quasar B1938+666 with a source redshift $z_\textrm{s}=0.881$ in the near-infrared band \cite{vegetti2012}. A similar result has been obtained for the lensed submillimeter galaxy SDP.81 with a source redshift $z_\textrm{s}=3.042$ \cite{inoue-minezaki2016,hezaveh2016a}. 

In this paper, we investigate a potential of power spectrum of astrometric shift (lensing shift power spectrum) to probe the properties of WDM and MDM. We first constrain the mass of WDM particles and the fraction of WDM in MDM models using a recent measurement owing to small scale structures (halos and troughs) in sightlines \cite{Inoue:2021dkv}.  
The use of astrometric shifts in a strongly lensed image is more advantageous than convergence if relative positions of de-lensed extended source images are measured precisely. Indeed, astrometric shifts can be directly measured from the relative positions of de-lensed extended source images that have two dimensional information. Therefore, we can constrain dark matter models with fewer number of lens samples compared to methods that use only flux ratios. It should be noted that convergence perturbation cannot be directly measured from flux ratios of lensed images of a point-like source since we need to model both the shear and convergence perturbations to explain anomalous flux ratios. Without the information of astrometric shifts, we have to resort to statistical methods that require a large number of samples. Moreover, we can easily observe astrometric shifts caused by intervening objects with subgalactic masses. Astrometric shifts can be significantly enhanced in the lens plane if the de-lensed images reside in the vicinity of caustics \cite{inoue2005b}. Typically, the amplitude of shift along the tangential direction of an Einstein ring is enhanced by a factor of $\gtrsim 10$ in the lens plane in cusp or fold caustics lenses \cite{inoue2005c}. Thus lensing shift power spectrum can be a powerful tool to constrain the WDM and MDM scenarios in which fluctuations of dark matter distribution on scales of $ \lesssim 10\,{\rm kpc} $ are suppressed. To our knowledge, lensing shift power spectrum has not been used so far to constrain dark matter models and hence this paper is the first one to utilize it as a cosmological dark matter probe.

The structure of this paper is the following. In the next section, we summarize how we calculate matter power spectra in WDM and MDM models. Then, in Section~\ref{sec:formalism}, we give our formalism to investigate the lensing power spectra, focusing on that of the astrometric shift to obtain constraints on the WDM mass and the fraction of WDM in the MDM model. In Section~\ref{sec:constraint}, we present constraints derived from a recent measurement of the lensing power spectrum, and expected ones attainable from the future observations. Then in the final section, we give the conclusion of this paper.

%%%%%%%%%%%%%%%%%%%%%%%%%%%%%%%%%%%%%%%%%%%%%%%%%%%%%%
%%%%%%%%%%%%%%%%%%%%%%%%%%%%%%%%%%%%%%%%%%%%%%%%%%%%%%
\section{Matter power spectra in warm dark matter and mixed dark matter models \label{sec:model}}
%%%%%%%%%%%%%%%%%%%%%%%%%%%%%%%%%%%%%%%%%%%%%%%%%%%%%%
%%%%%%%%%%%%%%%%%%%%%%%%%%%%%%%%%%%%%%%%%%%%%%%%%%%%%%

In this section, we briefly summarize the formulas to calculate non-linear matter power spectra in models with WDM and MDM. 

The prime effect of WDM particles is to suppress the structure in the Universe on scales below the free-streaming scale $\lambda_{\rm fs}$ which can be written as \cite{Kolb:1990vq}
\begin{equation}
\label{eq:FS}
  \lambda_{\rm fs} \simeq 0.2 ~{\rm Mpc} \, 
  \left( \frac{1~{\rm keV}}{m_{\rm WDM}}\right)
    g_\ast (t_{\rm NR})^{-1/2} g_{\ast s} (t_{\rm NR})^{1/3}
  \left[ 2 + \log \left( \frac{t_{\rm eq}}{t_{\rm NR}} \right)\right] \,,
\end{equation}
where $t_{\rm eq}$ and $t_{\rm NR}$ are the times at radiation-matter equality and when WDM particles become non-relativistic, respectively.  
$g_\ast(t_{\rm NR})$ and $g_{\ast s}(t_{\rm NR})$ are the effective number of degrees of freedom and its entropic counterpart at $t_{\rm NR}$. 
If all the WDM particles are thermally produced, the present energy density of WDM is given by   \cite{Kolb:1990vq} 
\begin{equation}
    \Omega_{\rm WDM} h^2 = 
    \left( \frac{T_{\rm WDM}}{T_\nu}\right)^3 
    \left( \frac{m_{\rm WDM}}{94~{\rm eV}}\right)^3  \,,
\end{equation}
where $T_{\rm WDM}$ and $T_\nu$ are the temperatures of WDM and neutrino, respectively.

To calculate the lensing power spectra, we need that for matter fluctuations on small scales, in which non-linear effects should be taken into account. In the following analysis, we adopt a fitting formula, constructed from N-body simulations to compute non-linear matter power spectra in WDM and MDM models assuming that baryonic effects are sufficiently small.  

\begin{figure}[ht]
\begin{minipage}[c]{0.5\hsize}
\begin{center}
\includegraphics[keepaspectratio, scale=0.75]{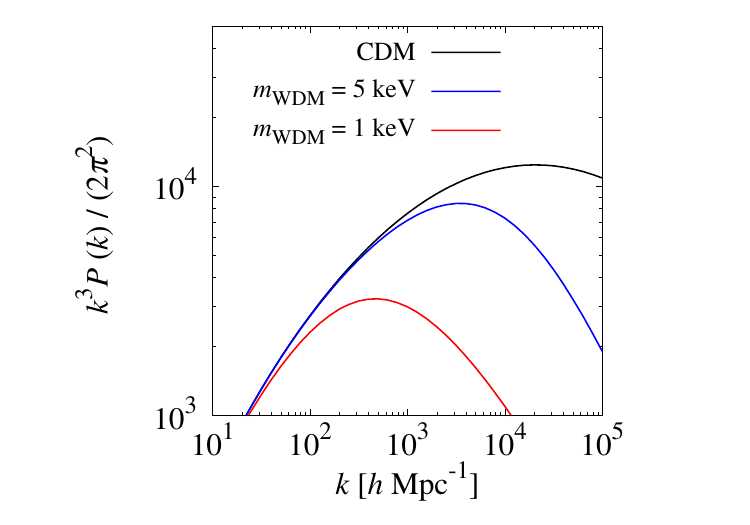}
\end{center}
\end{minipage}
\begin{minipage}[c]{0.5\hsize}
\begin{center}
\includegraphics[keepaspectratio, scale=0.75]{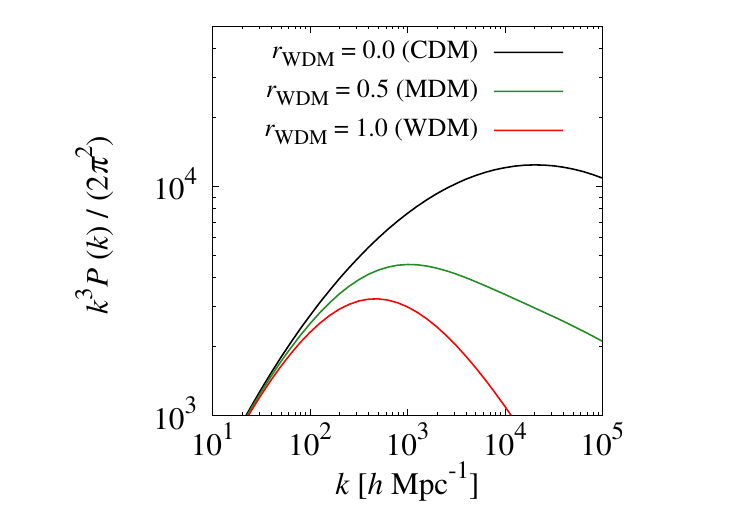}
\end{center}
\end{minipage}
      \caption{{\it (Left panel)}: Non-linear matter power spectra of pure WDM models at $z=0$.
      ({\it Right panel}): Non-linear power spectra of MDM models with $m_\mathrm{WDM}=1\,\mathrm{keV}$ at $z=0$. For comparison, we also plot the CDM case in both panels.
    \label{fig:P_NL}
    }
\end{figure}

Given a fitting function of matter power spectrum in the CDM model, which is denoted as  $P_{\rm CDM} (k,z)$, the corresponding WDM non-linear power spectrum $P_\mrm{WDM}(k,z)$ is given by \cite{inoue-etal2015}
\begin{align}
    \label{eq:P_WDM}
    \frac{P_\mrm{WDM}(k,z)}{P_\mrm{CDM}(k,z)} = \frac{1}{(1+k/k_d)^{0.7441}}\,,
\end{align}
with
\begin{align}
    \label{06110205}
    k_d(m_\mrm{WDM},z) = 388.8\,h\mrm{Mpc}^{-1}\,\biggl(\frac{m_\mrm{WDM}}{\mrm{1~keV}}\biggr)^{2.027} D^{1.583}(z)\,,
\end{align}
where $D(z)$ is the linear growth factor at $z$, which is  normalized as $D(0)=1$. A fitting function of non-linear matter power spectrum of CDM $P_{\rm CDM} (k,z)$ is given in~\ref{sec:app_nonlinear}. Non-linear matter spectra at $z=0$ in CDM and pure WDM models are shown in the left panel of Fig.~\ref{fig:P_NL} where we take the mass of WDM as $m_{\rm WDM} = 1~{\rm keV}$ and $5~{\rm keV}$. As the WDM mass decreases, the free-streaming scale becomes larger and small scale powers are more suppressed. 

Next we consider MDM models in which CDM and WDM coexist. To quantify the fraction of WDM, we define the fraction parameter as
\begin{align}
    \label{eq:r_WDM}
    r_\mrm{WDM} \equiv \frac{\Om_\mrm{WDM}h^2}{\Om_\mrm{CDM}h^2 +\Om_\mrm{WDM}h^2 } 
    = \frac{\Om_\mrm{WDM}h^2}{\Om_\mrm{DM}h^2}  \,,
\end{align}
where $\Om_\mrm{WDM}, \Om_\mrm{CDM}$ and $\Om_\mrm{DM}$ are the present WDM, CDM, and the total DM density parameters, respectively. The cases of $r_\mrm{WDM}=0$ and $1$ correspond to the standard CDM and the pure WDM models, respectively. Once we specify $m_{\rm WDM}$ and $r_{\rm WDM}$, one can obtain a non-linear matter power spectrum in the MDM model by adopting the following fitting formula obtained from N-body simulations \cite{Kamada:2016vsc}:
\begin{align}
    \label{eq:P_MDM}
    \frac{P_\mrm{MDM}(k,z)}{P_\mrm{CDM}(k,z)} = (1-f_\mrm{WDM}(r_{\rm WDM})) + \frac{f_\mrm{WDM}(r_{\rm WDM})}{(1+k/k_d^\prime)^{0.7441}}\,,
\end{align}
where $f_{\rm WDM}(r_{\rm WDM})$ is a function of $r_{\rm WDM}$, which is given as
\begin{align}
    \label{eq:f_warm}
    f_\mrm{WDM}(r_{\rm WDM}) = 1 - \exp\biggl(-\frac{1.551r_\mrm{WDM}^{0.5761}}{1-r_\mrm{WDM}^{1.263}}\biggr)\,,
\end{align}
and 
\begin{equation}
    \qquad k_d^\prime = \frac{k_d}{r_\mrm{WDM}^{5/6}} \,.
\end{equation}
Non-linear matter power spectra in MDM models at $z=0$ for $r_{\rm WDM} = 0,\ 0.5$ and $1$ are shown in the right panel of Fig.~\ref{fig:P_NL}. The mass of WDM is assumed as $m_{\rm WDM}=1~{\rm keV}$. The case with $r_{\rm WDM}=0$ corresponds to the CDM case.
As can be easily expected, for a smaller $r_{\rm WDM}$, the suppression of the non-linear matter power spectrum becomes less significant.  

%----------------------------------------------------------%
%----------------------------------------------------------%

By using the non-linear matter power spectra in WDM and MDM models described above, we can calculate lensing power spectra, which are going to be compared to observational results recently obtained in \cite{Inoue:2021dkv}, and expected data in future observations. In the next section, we describe how we can analyze them in WDM and MDM models given the WDM mass and the fraction of WDM.

%%%%%%%%%%%%%%%%%%%%%%%%%%%%%%%%%%%%%%%%%%%%%%%%%%%%%%
%%%%%%%%%%%%%%%%%%%%%%%%%%%%%%%%%%%%%%%%%%%%%%%%%%%%%%
\section{Lensing shift power spectrum \label{sec:formalism}}
%%%%%%%%%%%%%%%%%%%%%%%%%%%%%%%%%%%%%%%%%%%%%%%%%%%%%%
%%%%%%%%%%%%%%%%%%%%%%%%%%%%%%%%%%%%%%%%%%%%%%%%%%%%%%

In this section, we summarise our formalism to calculate the lensing power spectrum. In the following, we fix the cosmological parameters as  $H_0=67.36\,\mrm{km\,s^{-1}\,\mrm{Mpc}^{-1}}$, $\Om_{m}h^2=0.143$, $\Om_{b}h^2=0.02237$, and $\sigma_8 = 0.811$, which are derived from the Planck data \cite{Planck:2018vyg}.

First, let us suppose a lens system in which a point source at an angular position $\y$ is lensed with a deflection angle $\vec{\alpha}$. Then the angular position $\x$ 
of a lensed image is given by the lens equation, 
\begin{align}
\y=\x-\vec{\alpha}(\x).
\end{align}
Second, we consider a lens system in which a point source at a perturbed angular position $\tilde{\y}$ is lensed by
a primary lens that dominates the deflection of light with an angle $\vec{\alpha}$ and a secondary lens with an astrometric shift $\delta \vec{\alpha}$ with
$|\delta \vec{\alpha}|\ll |\vec{\alpha}|$ at the primary lens plane. Notice that an astrometric shift is given by the gradient of a projected gravitational potential perturbation $\delta \psi$ as $\delta \vec{\alpha}=\nabla \delta \psi$\footnote{
If a secondary lens (perturber) does not reside in the primary lens plane, we need to add a rotational (magnetic) component to the astrometric shift due to coupling between $\vec{\alpha}$ and $\delta \vec{\alpha}$ \cite{inoue2023}. In that case, $\delta \vec{\alpha}$ is interpreted as an electric (rotation free) component and the simple scaling relations between potential, shift, and convergence power spectra are broken. In the following, we neglect such a coupling effect as we expect that the contribution from halos in the vicinity of the source or observer are subdominant.
}.
Then the perturbed angular position $\tilde{\x}$ is given by  
\begin{align}
\tilde{\y}=\tilde{\x}-\vec{\alpha}(\tilde{\x})-\delta \vec{\alpha}(\tilde{\x}),
\end{align}
where  
\BEA
\tilde{\x}&=&\x+\delta \x,
\label{eq:gauge1}
\\
\tilde{\y}&=&\y+\delta \y ,
\label{eq:gauge2}
\EEA
with $\delta \x$ and $\delta \y$ being the perturbations of $\x$ and $\y$. Eqs.~(\ref{eq:gauge1}) and (\ref{eq:gauge2}) represent a relationship between the unperturbed and perturbed coordinates. Since the unperturbed coordinates $\x$ and $\y$ are not directly observable physical quantities, their choices are arbitrary. In other words, we have a two-dimensional gauge freedom for describing the shift of the images. 
In the `source plane gauge' in which $\delta \x=\boldsymbol{0}$, we have $\delta \y=-\delta \vec{\alpha}$\cite{Inoue:2021dkv}. Then the astrometric shift $\delta \vec{\alpha}$ with respect to the photon path in the macro (unperturbed) model is given by a shift $\delta \y$ in the source plane in the macro model.  

\begin{figure}[ht]
\begin{center}
    \includegraphics[clip,width=14.0cm]{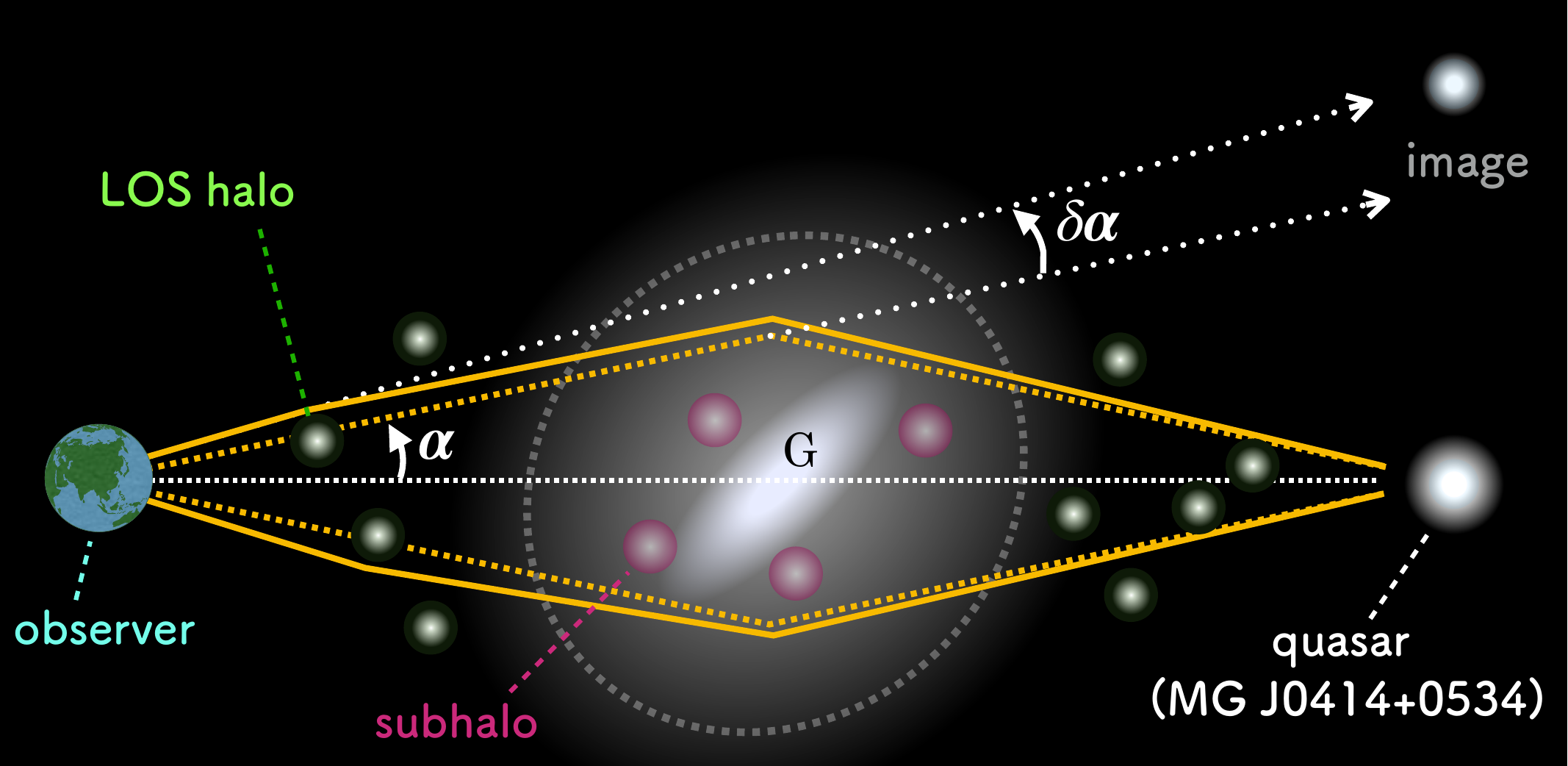}
    \caption{Illustration of our lens system. The yellow solid and dashed lines show the perturbed and non-perturbed paths, respectively.
    }
\label{fig:lens_system}
\end{center}
\end{figure}

An astrometric shift can be decomposed into various subcomponents (see Fig.~\ref{fig:lens_system}), such as contributions $\delta \vec{\alpha}_{\rm cluster/galaxy}$ from luminous galaxies and clusters, $\delta\vec{\alpha}_{\rm subhalo}$ from dark subhalos associated with the primary lens, and $\delta \vec{\alpha}_{\rm LOS}$ from line-of-sight (LOS) structures (dark halos and troughs) in the intergalactic space, which can be written as
\begin{align}
\delta \vec{\alpha}=\delta \vec{\alpha}_{\rm cluster/galaxy}+\delta\vec{\alpha}_{\rm subhalo}+\delta \vec{\alpha}_{\rm LOS}.
\end{align}  
As mentioned above, we assume that $|\delta\vec{\alpha}_{\rm subhalo}|$ and $|\delta \vec{\alpha}_{\rm LOS}|$ are sufficiently smaller than $|\vec{\alpha}|$ (weak lensing approximation), which is valid as long as the direction $\x$ is sufficiently far from the centers of these small-scale dark lens objects so that the strong lensing effect is negligible.

%----------------------------------------------------------%
%----------------------------------------------------------%

We assume that $\delta \vec{\alpha}_{\rm galaxy/cluster}$ can be subtracted by cutting off the scales below which this gives a sizable contribution, and its cut-off wavenumber is denoted as $k_{\rm min}$ which is described below. After the subtraction, we have
\begin{align}
\delta \vec{\alpha} \approx \delta\vec{\alpha}_{\rm subhalo}+\delta \vec{\alpha}_{\rm LOS}.
\end{align}   
The 2-point correlation function $\langle \delta \vec{\alpha}(\x)\cdot \delta \vec{\alpha}(\x') \rangle$ describes a weak lensing effect owing to the subhalos and the LOS structures. Moving to the Fourier space, the lensing shift power spectrum defined on a region with a solid angle of $A$ that includes lensed arcs is given by
\begin{align}
 P_{\alpha}(l)\equiv \frac{A}{(2 \pi)^2}\biggl\langle\Bigl|\hat{\delta} \alpha(\vec{l})\Bigr|^2 \biggr\rangle, 
\end{align}
where $l=|\vec{l}|$ with $\vec{l}$ being an angular wave vector which can be written in terms of an angular scale $\delta \theta$ as \cite{Inoue:2021dkv}
\begin{equation}
    l = \frac{1.296 \times 10^6}{(\delta \theta / 1^{''})} \,.
\end{equation}
$\hat{\delta} \alpha(\vec{l})$ is the Fourier transform of $\delta \alpha(\vec{x})$ in the comoving coordinates which is given as
\begin{equation}
\hat{\delta} \alpha(\vec{l}) = \int d^2 x \, \delta \alpha (\vec{x}) \,  e^{-i \vec{l} \cdot \vec{x}} \,.
\end{equation}
The `dimensionless' power, which represents the amplitude of astrometric shift at an angular scale of $\sim l^{-1}$ per logarithmic interval of $l$ is defined as
\begin{align}
\label{eq:Delta_def}
 \Delta^2_\alpha (l)\equiv 2 \pi l^2 P_{\alpha}(l).
\end{align}

Considering the LOS contribution, the two-point correlation function of $\de \alpha$ is approximately given by \cite{Bartelmann:1999yn, Inoue:2012px}
\begin{align}
    \label{eq:xi_shift}
    \xi_{\al} (\theta) &\equiv \langle \de\al(\bm{0})\,\de\al(\bm{\theta}) \rangle 
    = 4\int^{r_S}_{0} dr\ \biggl(\frac{Q(r)}{g(r)}\biggr)^2\, 
     \int^{k_\mrm{max}}_{k_{\rm min}} \frac{dk}{2\pi k}\   P(k;r)\, J_0\bigl(kg(r)\,\theta\bigr)\,,
\end{align}
where 
\begin{align}
    Q(r) = \frac{3 H_0^2\Om_{m,0}}{2c^2}\,\frac{r(r-r_S)}{r_S}\,\bigl[1+z(r)\bigr]\,,
\end{align}
and 
\begin{align}           
    g(r) = \left\{
\renewcommand{\arraystretch}{2.0}
\begin{array}{ll}
r &\quad(r<r_L)\\
\displaystyle{\frac{r_L(r_S-r)}{r_S-r_L}} &\quad(r\geq r_L)\,,
\end{array}\right.
\end{align}
with $r_S$ and $r_L$ being the comoving distance to the source and the primary lens, respectively. $g(r)$ describes an unperturbed light path, $J_0$ is the zero-order Bessel function, and $P(k;r)$ is the non-linear matter power spectrum, which is given by $P_{\rm WDM}$ or $P_{\rm MDM}$ (Eq.~\eqref{eq:P_WDM} or \eqref{eq:P_MDM}) derived from fitting the outcomes of N-body simulations for the pure WDM or mixed DM cases. $k_{\rm min}$ corresponds to the wavenumber above which  $\delta \vec{\alpha}_{\rm cluster/galaxy}$ can be neglected, and describes the possible sizes and positions of LOS halos.
Actually, since $k_\mrm{min}$ is difficult to determine, here we take a conservative approach with regard to the treatment of $k_\mrm{min}$: we choose $k_\mrm{min}$ such that the data gives a best fit when $k_\mrm{min}$ is varied.  
In other words, $k_\mrm{min}$ is treated as a model parameter in our analysis.

This treatment is equivalent to adding the sharp $k$-space filter in the integrand in Eq.~(\ref{eq:xi_shift}) \cite{inoue-takahashi2012, takahashi-inoue2014}. We adopt the upper cut-off $k_\mrm{max}$ in the $k$-integral as $k_\mrm{max}=10^5\,h\mrm{Mpc}^{-1}$, which corresponds to the observed small scale structures in the quasar host galaxy \cite{Inoue:2021dkv}. Note that we assumed that the emitted light propagates on the surface of a combined pair of cones in the comoving space, which is a good approximation in the vicinity of a lensed arc \cite{Takahashi:2012em}.

By using the two-point correlation function, the lensing shift power spectrum can be written as
\begin{align}
\label{eq:PS_shift}
P_\al(l) &= 2\pi \int d\theta\ \theta \xi_\al(\theta)\,J_0(l\theta)\nn\\
&= 4\int^{r_S}_0 dr\ \biggl( \frac{Q(r)}{g(r)} \biggr)^2\, \int^{k_{\rm max}}_{k_\mrm{min}} \frac{dk}{k}\, P(k;r)\, \frac{\de_D\bigl(k-l/g(r)\bigr)}{l g(r)} \nn  \\[8pt]
&= \left\{
\renewcommand{\arraystretch}{2.0}
\begin{array}{ll}
\displaystyle{\frac{4}{l^2}\,\int^{r_S}_0 dr\, \biggl( \frac{Q(r)}{g(r)} \biggr)^2\,P\biggl(\frac{l}{g(r)};r\biggr)}\, &\quad\bigl(l/r_L \leq k_\mathrm{min}\bigr) \\
0\, &\quad(\mrm{otherwise})\,.
\end{array}\right.
\end{align}
In the following analysis, we use $\Delta_\al(l)$, defined in Eq.~\eqref{eq:Delta_def}, instead of the power spectrum $P_\alpha(l)$.

Here we briefly mention the relation between the lensing shift and convergence power spectra.  
If a perturbing object resides at the primary lens plane, the astrometric shift $\de\bm\al(\bm\theta)$  owing to the perturber is related to a convergence perturbation $\delta \kappa (\bm \theta)$ as
\begin{align}
    \label{eq:convergence}
\delta \ka(\bm\theta) \ =\ \frac{1}{2}\,\nabla_{\bm\theta}\cd \delta \bm\al(\bm\theta)\,,
\end{align}
where $\nabla_{\bm\theta}$ denotes the 2-dimensional gradient operator (see, e.g., \cite{Bartelmann:1999yn}).
Then, the convergence power $\Delta_\ka$ is proportional to the shift power $\Delta_\al$  as
\begin{align}
    \label{eq:PS_a-p}
    \Delta_\ka(l) = \frac{l}{2}\,\Delta_\al(l)\,.
\end{align}
If a perturbing object does not reside at the primary lens plane, the relation Eq.~\eqref{eq:PS_a-p} no longer holds and another term that is linear to the LOS distance parameter appears in the right hand side of Eq.~\eqref{eq:PS_a-p} due to coupling between $\bm \alpha$ and $\delta \bm \alpha$ \cite{inoue2023}.

In our analysis, we constrain the WDM mass and the fraction of WDM in the MDM model using a lensing shift power spectrum, which was measured using ALMA (Atacama Large Milllimeter/submillimeter Array) observations of the quadruply lensed quasar MG J0414+0534 \cite{Inoue:2021dkv}. The redshifts of the source and the primary lens (early type galaxy) of MG\,J0414+0534 are $z_S=2.639$ \cite{lawrence1995} and $z_L=0.9584$ \cite{tonry1999}, respectively. These redshifts are adopted in the our analysis. The lensing shift power spectrum was obtained by fitting discrete Fourier modes of potential perturbation that satisfy the Dirichlet boundary condition on the sides of a square with a sidelength of $L$ to the observed data.  The lens system consists of a massive early type galaxy G, which acts as a primary lens,
and an object X, which is possibly a less massive galaxy, and an object Y, which is possibly a dusty dwarf galaxy. The gravitational effects of object X and object Y were subtracted using a smooth potential for each object. In the following analysis, we use only the lensing shift power at the angular wave number of $l \sim 1.3 \times 10^6$ with $L=\asec{3}{6}$, which roughly corresponds to the effective Einstein radius $\asec{1}{1}$ of galaxy~G.  We expect that the measured powers on angular scales larger than the Einstein radius are sensitive to the Dirichlet boundary condition because of the proximity in the scale. Moreover, lack of information on the radial direction in the primary lens plane due to the small size of source would lead to a suppression of powers on large angular scales, which is the reason why we do not use the data on larger scale, but only use the data at $l \sim 1.3 \times 10^6$.

%----------------------------------------------------------%
\renewcommand{\arraystretch}{1.5}
\begin{table}[ht]
  \caption{\label{table:observed_L36} 
  Observed lensing power spectrum $\Delta_\alpha\,[\mrm{arcsec}]$ at $l \sim 1.3 \times 10^6$ with $L=\asec{3}{6}$ \cite{Inoue:2021dkv}. $\sigma_{\alpha, \rm obs}, \sigma_{\al, \rm cv}$ and $\sigma_{\al, \rm msd}$ are 1$\sigma$ errors originating from the measurement quoted in \cite{Inoue:2021dkv}, the cosmic variance and the mass-sheet degeneracy, respectively.
  } \vspace{3mm}
  \centering
  \begin{tabular}{|c|cccc|}
    \hline
    $l$ & $\Delta_\al$ & $\sigma_{\alpha, \rm obs}$ & $\sigma_{\al, \rm cv}$ & $\sigma_{\al, \rm msd}$\\
     \hline \hline 
    $1.3^{+0.08}_{-0.08}\times10^6$ & 0.0081 & $\pm$ 0.0006  & $\pm$0.00153 & $\pm$0.00081 \\
    \hline
  \end{tabular}
\end{table}
%----------------------------------------------------------%

Table~\ref{table:observed_L36} shows the measured angular wave number, the mean value of the lensing shift power, and its 1$\sigma$ uncertainty $\sigma_{\alpha, \rm obs}$ \cite{Inoue:2021dkv}. We also consider two other types of error that contribute to the uncertainty, which are not included in the quoted error in \cite{Inoue:2021dkv}.  One is the cosmic variance error $\sigma_{\al, \rm cv}$ due to a limited number of the measured discrete Fourier modes $N_\mrm{mode}$, which is given by $\De_\al/(2N^{1/2}_\mrm{mode})$ where $N_\mrm{mode}=7$ for the bin of $l\sim1.3\times10^6$ \cite{Inoue:2021dkv}. The other one is the error due to the mass sheet degeneracy~\cite{1985falco}, which is denoted as $\sigma_{\al, \rm msd}$ and yields uncertainty in the Hubble constant obtained from time delay. We assume a conservative $10\,\%$ error due to the ambiguity in the mass profile of the primary lens of MG\,J0414+0534 \cite{2013schneider}. Then, we include $10\,\%$ of the shift power spectrum for CDM as the error since the result of \cite{Inoue:2021dkv} is consistent with the CDM model.

%----------------------------------------------------------%
\renewcommand{\arraystretch}{1.0}
\begin{figure}[t]
         \begin{minipage}[c]{0.5\hsize}
        \begin{center}
          \includegraphics[keepaspectratio, scale=0.7]{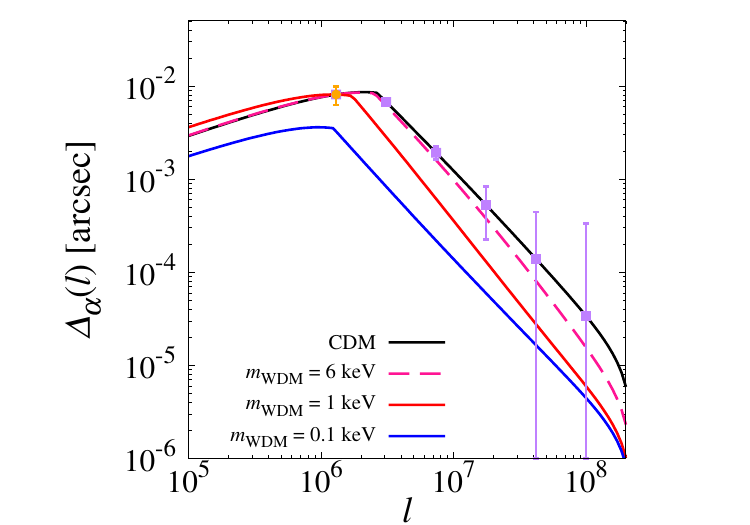}
          \end{center}
          \end{minipage}
         \begin{minipage}[c]{0.5\hsize}
          \begin{center}
          \includegraphics[keepaspectratio, scale=0.7]{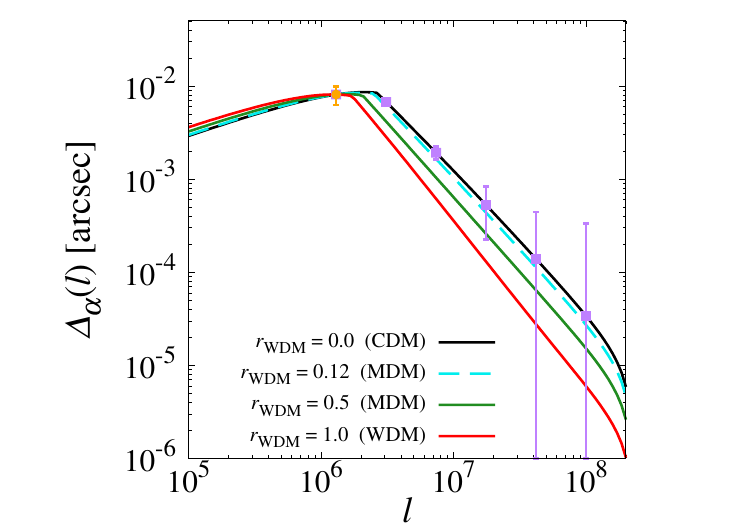}
           \end{center}
          \end{minipage}
 \caption{The lensing power spectra for WDM (left) and MDM with $m_\mrm{WDM}=1\,\mrm{keV}$ (right) toward $z_S=2.639$.  Orange circle and the error bar show the measured value for $L=3.6\,\mrm{arcsec}$ in \cite{Inoue:2021dkv}. Purple squares represent mock data (for the case of $N_L=30$ with $N_L$ the number of lens system; for the detail, see the next section) expected from future observations, which have the same value as the one of the power spectrum for CDM. $k_{\rm min}$ is set such that the lensing shift powers are best-fitted to the measured one for the parameter set: $k_\mrm{min}\simeq1190\,h\mrm{Mpc}^{-1}$ (CDM), $k_\mrm{min}\simeq773\,h\mrm{Mpc}^{-1}$ ($m_{\rm WDM} = 1\,\mrm{keV}$), $k_\mrm{min}\simeq585\,h\mrm{Mpc}^{-1}$ ($m_{\rm WDM} = 10.1\,\mrm{keV}$), $k_\mrm{min}\simeq925\,h\mrm{Mpc}^{-1}$ ($r_\mrm{WDM}=0.5$), and $k_\mrm{min}\simeq1121\,h\mrm{Mpc}^{-1}$ ($r_\mrm{WDM}=0.12$). $k_\mrm{min}$ for  each dashed line is obtained by fitting the model to the mock data. 
 }
 \label{fig:Delta}
\end{figure}
%----------------------------------------------------------%

In the following analysis, we neglect contribution from subhalos that reside in the halo of the primary lens,  and directly compare the theoretically computed lensing shift power caused by the LOS (i.e., Eq.~(\ref{eq:PS_shift})) with
the measured power given in Table~\ref{table:observed_L36}. In CDM models, a recent theoretical study show that subhalos in the primary lens of MG\,J0414+0534 are expected to contribute just one third of the total magnification effect~\cite{inoue2016}. In WDM or MDM models, the ratios of the subhalo to LOS contributions are not known. However, we expect that the subhalo contribution is subdominant. The reason is as follows: The lensing effect of subhalos in WDM or MDM models would be much weaker than that in the CDM model due to the suppression of small scale fluctuations. It is likely that suppression of LOS structures in WDM or MDM models are less stringent than that of subhalos, because subhalos should be formed at a redshift higher than $z_L$ at which suppression is much stronger than lower redshifts $z<z_L$. Thus we expect that the subhalo contribution is subdominant in WDM or MDM models.

%----------------------------------------------------------%
\begin{figure}[t]
\begin{center}
    \includegraphics[clip,width=12.5cm]{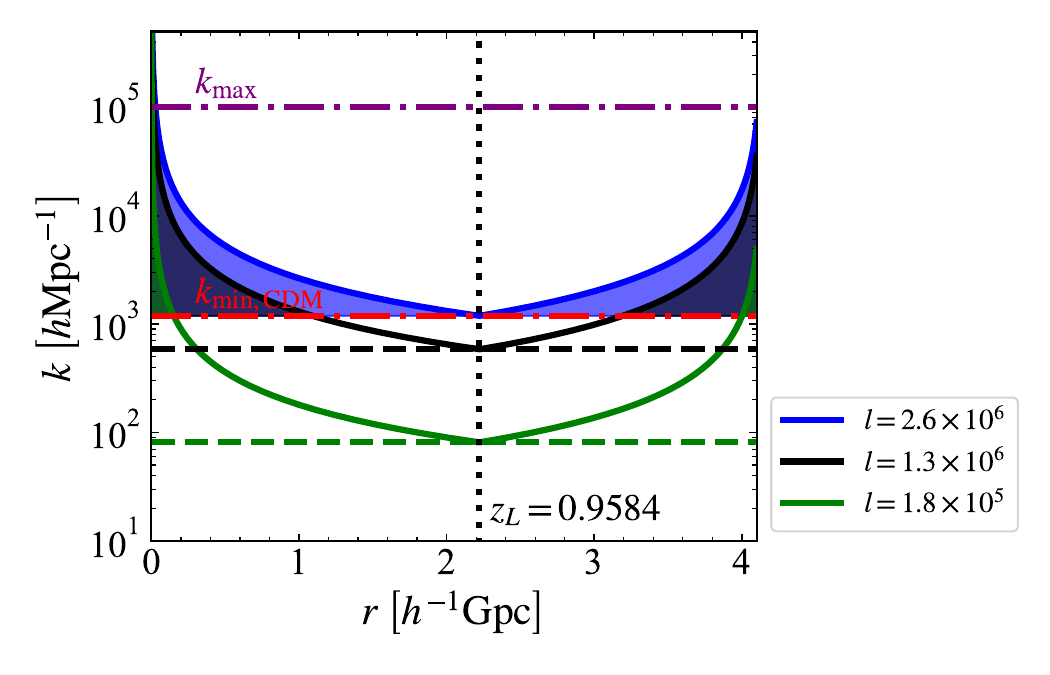}
    \caption{Dependence of the wave number on the comoving distance $r$, which is determined by $k=l/g(r)$. The dashed lines represent to the value of $k_L$  corresponding the comoving distance to $z_L$ for each $l$ at which $k$ takes the minimum value. The red dash-dotted line corresponds to $ k_\mrm{min}\simeq1190\,h\mrm{Mpc}^{-1}$ which gives the best-fit shift power spectrum for the CDM case. The purple dash-dotted line is for $k_\mrm{max}=10^5\,h\mrm{Mpc}^{-1}$, which is fixed in our analysis.  The shaded region indicates the range that contributes to the shift power spectrum for a given $l$.}
\label{fig:r-k}
  \end{center}
\end{figure}

Fig.~\ref{fig:Delta} shows the power spectra of astrometric shift for several values of $m_{\rm WDM}$ and $r_{\rm WDM}$,  as well as the measured one from \cite{Inoue:2021dkv} with the $1\sigma$ total error $\sigma_{\rm total}$ which is evaluated as 
\begin{equation}
\sigma_{\rm total} 
= \sqrt{\sigma_{\alpha, \rm obs}^2 + \sigma_{\al, \rm cv}^2 + \sigma_{\al, \rm msd}^2}\,.
\end{equation}
We also show the spectra in the CDM model for comparison. 
As seen in Fig.~\ref{fig:Delta}, as $m_{\rm WDM}$ ($r_{\rm WDM}$) decreases (increases), the amplitude of the power spectrum decreases, from which one can see how stringent constraints on $m_{\rm WDM}$ and $r_{\rm WDM}$ can be obtained by comparing with the error depicted. The purple points and error bars are the mock data expected in future observations, which we will give some detail in the next section. 
We also note that the lensing shift power spectrum for the CDM case takes a maximum value at $l\sim 2.6 \times 10^6$, which corresponds to the scale at which $k_{\rm min}$ is given by $k_L = l / g(r_L)$ with $k_L$ and $r_L$ respectively being the wavenumber and comoving distance corresponding to $z_L$ (see Eq.~\eqref{eq:PS_shift}).  Indeed the contribution from the LOS halos becomes the largest when $k_L = k_{\rm min}$. Fig.~\ref{fig:r-k} illustrates that, for the shift power spectrum at $l$ smaller than $2.6 \times 10^6$ (in the figure, the cases with $l=  1.3 \times 10^6$ and $1.8 \times 10^6$ are shown), the range of scale $k$ is restricted since the wavenumber should satisfy $k > l/g(r)$.
For example, in the CDM case, we find that only LOS halos residing at  $r/(h^{-1}\mrm{Gpc}) \lesssim 1$ or $3\gtrsim r/(h^{-1}\mrm{Gpc})$ contribute to the shift power spectrum at $l=1.3\times10^6$. The restriction on the location of LOS halos is much severer for the power at $l=1.8\times10^5$. On the other hand, at $l=2.6\times10^6$, there is no restriction on the location of LOS halos that can contribute to the integral. This is the reason why the lower cutoff $k_\mrm{min}$ affects the the lensing shift power spectrum.

%%%%%%%%%%%%%%%%%%%%%%%%%%%%%%%%%%%%%%%%%%%%%%%%%%%%%%%%%%%
\section{Constraints on WDM and MDM \label{sec:constraint}}
%%%%%%%%%%%%%%%%%%%%%%%%%%%%%%%%%%%%%%%%%%%%%%%%%%%%%%%%%%%

Now, in this section, we discuss constraints on the mass and the fraction of WDM from a recent measurement \cite{Inoue:2021dkv} and expected data in future observations. In order to obtain constraints on the parameters $m_{\rm WDM}$ in the pure WDM model, and  $m_{\rm WDM}$ and $r_{\rm WDM}$ in the MDM model, we evaluate the $\chi^2$ defined as 
\begin{align}
    \label{eq:chi2}
    \chi^2  = \sum_i\, \frac{\bigl(\De^{\rm th}_\al - \De_\al^\mrm{obs}\bigr)^2}{\sigma_{\alpha, \rm obs}^2 + \sigma_{\al, \rm cv}^2 + \sigma_{\al, \rm msd}^2 }\Biggr|_{l=l_i}\,,
\end{align}
where $\De^{\rm th}_\al $ is a theoretically predicted shift power spectrum in the WDM and MDM models and $\De^{\rm obs}_\al$ is the observed one. $\sigma_{\al, \rm obs}$, $\sigma_{\al, \rm cv}$ and $\sigma_{\al, \rm msd}$ are uncertainties described in the previous section.  The values of $ \De_\al^\mrm{obs}$, $\sigma_{\al, \rm obs}$, $\sigma_{\al, {\rm cv}}$  and $\sigma_{\al, {\rm mds}}$  are given in Table~\ref{table:observed_L36}.
By calculating $\chi^2$,  we evaluate the bounds on $m_{\rm WDM}$ for the pure WDM model, and $m_{\rm WDM}$ and $r_{\rm WDM}$ for the MDM model. First we present our results for the pure WDM case, then those for the MDM case follow.

%%%%%%%%%%%%%%%%%%%%%%%%%%%%
%%%%%%%%%%%%%%%%%%%%%%%%%%%%
\subsection{Pure WDM case}

%----------------------------------------------------------%
\begin{figure}[ht]
\begin{center}
    \includegraphics[clip,width=10.0cm]{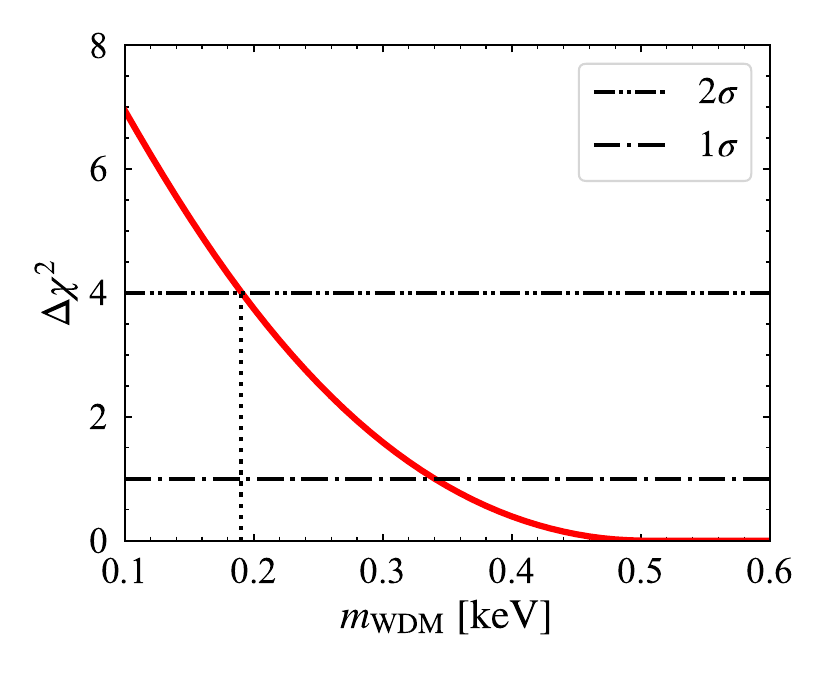}
    \caption{Plot of $\Delta\chi^2\equiv\chi^2-\chi^2_\mrm{min}$ for the WDM model. The dot-dashed line shows 1$\si$ error, while the dot-dot-dashed one indicates $2\si$ error.
\label{fig:Delta_chi2_pure}
}
  \end{center}
\end{figure}
%----------------------------------------------------------%

%----------------------------------------------------------%
\begin{figure}[ht]
\begin{center}
    \includegraphics[clip,width=10.0cm]{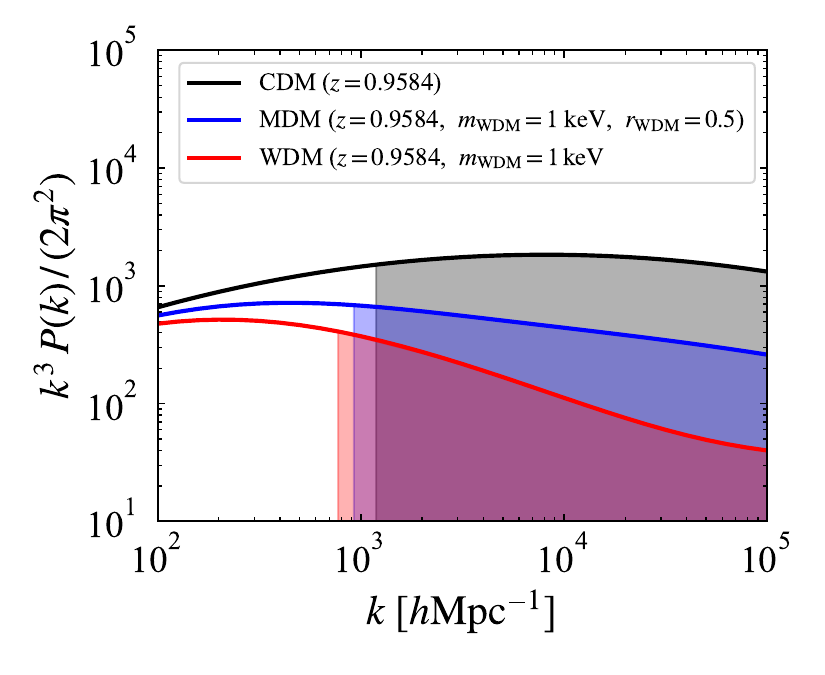}
    \caption{Non-linear power spectra at $z=0.9584$ which are used to calculate the lensing shift power spectra for CDM and non-CDM models. The shaded regions represent the range of the power spectra contributing to the integral~\eqref{eq:PS_shift}, i.e., $k_{\rm min} < k < k_{\rm max}$.}
\label{fig:P-kmin}
  \end{center}
\end{figure}
%----------------------------------------------------------%

Fig.~\ref{fig:Delta_chi2_pure} shows $\Delta\chi^2$ as a function of $m_{\rm WDM}$, from which we can read off the 2$\sigma$ bound on the WDM mass as 
\begin{equation}
    m_{\rm WDM} \gtrsim 0.19~{\rm keV} \qquad (95~\%~{\rm C.L.})\,.
\end{equation}
Although this bound is not as strong as other existing constraints, it should be noted that the above limit is obtained only with a single lens system. Once we observe more lens systems, which is expected to be the case in the near future, we would be able to obtain much tighter bound. We discuss attainable constraints from expected future data in Sec.~\ref{sec:future}.

In Fig.~\ref{fig:P-kmin}, we show non-linear power spectra for the cases with the pure WDM, MDM and CDM cases (the parameters assumed in each case are shown in the legend of the figure), and the range of scales $k_{\rm min}< k < k_{\rm max}$ are depicted with shaded colors, which contribute to the lensing shift power spectrum. As seen from the figure, $k_\mrm{min}$ in the CDM model is 
larger than the non-CDM models, which implies that the distance between a LOS halo and the primary lens (galaxy G) in the CDM model is expected to be larger than those in the non-CDM models. Therefore, if the redshifts of LOS halos are measured, the property might be used to check consistency with the adopted model.

%%%%%%%%%%%%%%%%%%%%%%%%%%%%
%%%%%%%%%%%%%%%%%%%%%%%%%%%%
\subsection{Mixed DM case}

%----------------------------------------------------------%
\begin{figure}[ht]
\begin{center}
    \includegraphics[clip,width=10.0cm]{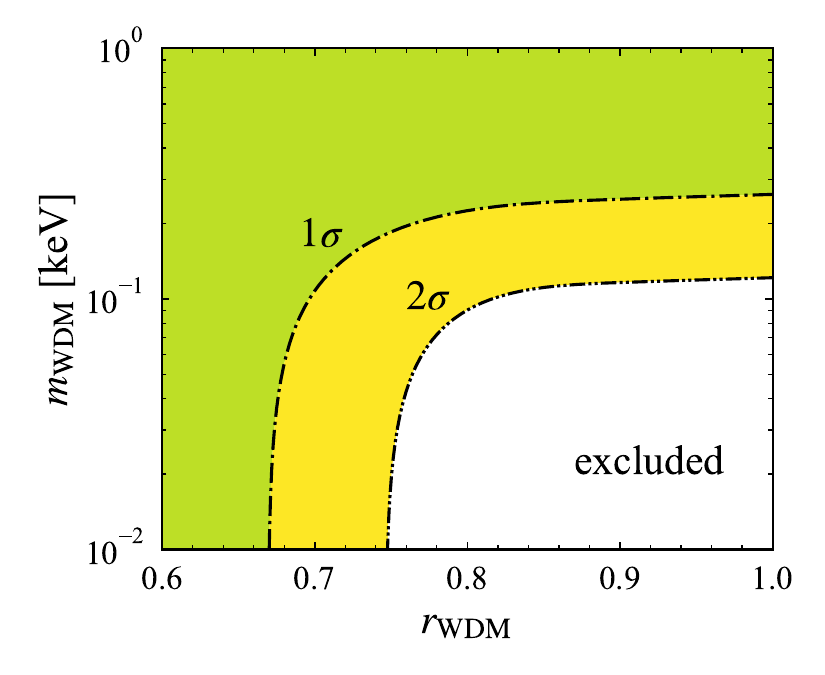}
    \caption{\label{fig:Delta_chi2_mix}
    1$\sigma$ and 2$\sigma$ constraints in the $r_{\rm WDM}$--$m_{\rm WDM}$ plane.
    }
  \end{center}
\end{figure}
%----------------------------------------------------------%

Next we discuss the case of MDM, in which WDM and CDM particles coexist. In this case, we have two parameters to characterize the model: $m_{\rm WDM}$ and $r_{\rm WDM}$.

In Fig.~\ref{fig:Delta_chi2_mix}, $1\sigma$ and $2\sigma$ allowed regions are shown in the $m_{\rm WDM}$--$r_{\rm WDM}$ plane. Needless to say, even with the same $m_{\rm WDM}$, when the fraction of WDM is small, the suppression of the power spectra becomes milder as seen from Fig.~\ref{fig:P_NL} (see, also Fig.~\ref{fig:P-kmin}). 
The fraction of WDM is constrained, for example, in the case of $m_{\rm WDM} = 0.1~{\rm keV}$, as 
$r_{\rm WDM} \lesssim 0.72$. As in the case of the pure WDM model, the current constraint is not so severe, however, it can be much tighter in the near future, which will be also discussed in the next section.

%%%%%%%%%%%%%%%%%%%%%%%%%%%%
%%%%%%%%%%%%%%%%%%%%%%%%%%%%
\subsection{Future expected constraints \label{sec:future}} 

In the analysis of the previous sections, we have used the measured lensing shift power spectrum derived from a single lens system alone. If more similar lens systems are added to evaluate the lensing power spectrum, the constraints would be significantly improved. On-going and planned observations with ALMA \cite{ALMA} and JWST \cite{JWST} will surely improve the constraint on dark matter models. In the near future, next-generation ground telescopes such as TMT \cite{TMT}, ELT \cite{ELT}, and ngVLA \cite{ngVLA} will become very powerful tools by measuring the lensing shift power spectrum with better accuracy on smaller scales.

To investigate to what extent constraints will become more stringent in future observations, we assume a measurement of $\Delta_\alpha$ with five bins for the range of $3.1\leq l/10^6\leq8.0$, which are equally spaced on a logarithmic scale (see also Fig.~\ref{fig:Delta}). Indeed, such scales could be observed by ngVLA \cite{ngVLA} and VLBA \cite{VLBA}. Each center value is given by Eq.~\eqref{eq:PS_shift} for the CDM model.
We assign the uncertainties of the mock measurement as follows. Although the lensing power spectrum would be more precisely measured in the future observations, we conservatively assume that the lensing shift power spectrum per lens system is the same as the one in the power for MG\,J0414+0534 and it does not depend on angular scales, and the systematic errors are negligible. However, 
given that the number of strong lens system is increased from one to $N_L$, the uncertainties
decrease as $\sigma \propto 1 /\sqrt{N_L}$. Here we consider the cases with (i) $N_L=30$ and (ii) $N_L=100$, which would be achievable in about a few to 10 years, respectively \cite{canameras2021, shu2022}. 
To evaluate the future projected constraints on $m_{\rm WDM}$ and $r_{\rm WDM}$, we replace the uncertainties for the lensing power spectrum as
\begin{eqnarray}\label{eq:future_error}
\sigma_{\al, {\rm obs}} \rightarrow  \frac{\sigma_{\al, {\rm obs}}}{\sqrt{N_L}} \,,\qquad 
\sigma_{\al, {\rm cv}} \rightarrow  \frac{\sigma_{\al, {\rm cv}}}{\sqrt{N_L}} \,, \qquad 
\sigma_{\al, {\rm msd}} \rightarrow  \frac{\sigma_{\al, {\rm msd}}}{\sqrt{N_L}} \,,
\end{eqnarray}
where $\sigma_{\al, {\rm obs}}$, $\sigma_{\al, {\rm cv}}$, and $\sigma_{\al, {\rm msd}}$ correspond to the errors given in Table~\ref{table:observed_L36}, respectively. 
We assume that the correlations between the errors in different bins are negligible.

%----------------------------------------------------------%
\renewcommand{\arraystretch}{1.5}
\begin{table}[ht]
  \caption{\label{tab:future_const_mWDM}
  Projected 2$\sigma$ constraints on $m_{\rm WDM}$ expected in future observations for the cases with the fraction of WDM fixed as $r_{\rm WDM} = 1$, $0.75$ and $0.5$.
  } \vspace{3mm}
  \centering
  \begin{tabular}{|r|ccc|}
    \hline
    $N_L$ &  $ r_{\rm WDM}=1$ (pure WDM)  & $r_{\rm WDM}=0.75$ (mixed)  & $r_{\rm WDM}=0.5$ (mixed) \\
     \hline  
    $30$  & $\gtrsim$ 6.3 keV & $\gtrsim$ 4.8 keV & $\gtrsim$ 3.6 keV \\
    $100$ & $\gtrsim$ 9.0 keV & $\gtrsim$ 7.0 keV & $\gtrsim$ 5.3 keV \\
    \hline
  \end{tabular}
\end{table}
%----------------------------------------------------------%
%----------------------------------------------------------%
\begin{table}[ht]
  \caption{\label{tab:future_const_rWDM}
  Projected 2$\sigma$ constraints on $r_{\rm WDM}$ expected in future observations for the cases with the WDM masss fixed as $m_{\rm WDM} = 2~{\rm keV}$, $5~{\rm keV}$ and $7~{\rm keV}$.
  } \vspace{3mm}
  \centering
  \begin{tabular}{|r|ccc|}
    \hline
    $N_L$ & $m_{\rm WDM}=2~{\rm keV}$  & $m_{\rm WDM}=5~{\rm keV}$  & $m_{\rm WDM}=7~{\rm keV}$\\
     \hline  
    $30$  & $\lesssim$ 0.25  & $\lesssim$ 0.80  & $\leq$ 1  \\
    $100$ & $\lesssim$ 0.14  & $\lesssim$ 0.46 & $\lesssim$ 0.74  \\
    \hline
  \end{tabular}
\end{table}
\renewcommand{\arraystretch}{1.0}
%----------------------------------------------------------%

In Table~\ref{tab:future_const_mWDM}, we show future projected 2$\sigma$ constraints on $m_{\rm WDM}$ for $r_{\rm WDM}=1$, $0.75$ and $0.5$ where the first one corresponds to the pure WDM case and the latter two to the MDM one.  In Table~\ref{tab:future_const_rWDM}, the projected 2$\sigma$ constraints on $r_{\rm WDM}$ are listed.  As seen from the tables, we can obtain more stringent bounds on the WDM mass and the fraction of WDM once more lens systems to evaluate the power spectrum are observed, which shows that our method using the lensing shift power spectrum would be very promising to constrain the pure WDM and MDM models although the current constraint is not yet so strong.

%%%%%%%%%%%%%%%%%%%%%%%%%%%%%%%%%%%%%%%%%%%%%%%%%%%%%%%%%%%
\section{Conclusion and discussion  \label{sec:conclusion}}
%%%%%%%%%%%%%%%%%%%%%%%%%%%%%%%%%%%%%%%%%%%%%%%%%%%%%%%%%%%

In this paper, we have investigated the constraints on WDM and MDM models by using the lensing shift power spectrum obtained in \cite{Inoue:2021dkv}. To our knowledge, our analysis represents the inaugural attempt to utilize the lensing shift power spectrum for constraining cosmological models. Notably, our study is pioneering in deriving constraints on WDM and MDM models through this tool.

While it is in principle feasible to combine information from lensing power spectra for convergence and potential, as also acquired in \cite{Inoue:2021dkv}, we adopt a conservative approach in our investigation. We exclusively rely on the directly measurable astrometric shift power spectrum, which is less susceptible to systematic errors inherent in observations.

By leveraging the lensing shift power spectrum obtained from a single lens system, reported in \cite{Inoue:2021dkv}, we establish a constraint on the WDM mass as
\begin{equation}
   m_{\rm WDM} \gtrsim 0.19~{\rm keV} \qquad (95~\%~{\rm C.L.})\,,
\end{equation}
for the scenario where WDM constitutes the entire dark matter content. In the MDM models where WDM coexists with CDM, we introduce the parameter $r_{\rm WDM}$ characterizing the fraction of WDM in the total DM. The constraints on $r_{\rm WDM}$ are contingent upon the WDM mass and are illustrated in Fig.~\ref{fig:Delta_chi2_mix}.

The above analysis is based on a single quasar lens system. While the current constraints on WDM and MDM from this single system may not rival the stringency of other methodologies, the potential for much tighter constraints arises with the availability of additional lens systems, contingent on well-controlled systematic error mitigation.

Anticipating forthcoming observations, we project future bounds on $m_{\rm WDM}$ and $r_{\rm WDM}$, as shown in Tables~\ref{tab:future_const_mWDM} and~\ref{tab:future_const_rWDM}. In the pure WDM case, the expected constraints on the WDM mass $m_{\rm WDM}$,
would be
\begin{equation}
   m_{\rm WDM} \gtrsim 6.3~{\rm keV} \qquad (95~\%~{\rm C.L.})\,,
\end{equation}
for $N_L=30$, and 
\begin{equation}
   m_{\rm WDM} \gtrsim 9.0~{\rm keV} \qquad (95~\%~{\rm C.L.})\,,
\end{equation}
for $N_L=100$, both of which are obtained by assuming the CDM model as a fiducial one. Ongoing observations with ALMA and JWST, coupled with future possibilities with TMT, ELT, and ngVLA within the next decade, 
would bring the lensing shift power spectrum a robust tool for exploring a range of non-CDM models.

To attain more stringent constraints on non-CDM models, refinement of our analysis methodology is imperative. In our current approach, we neglected the coupling effect between the deflection $\vec{\alpha}$ from the primary lens and that from the secondary lens $\delta \vec{\alpha}$. While this simplification suffices for the present analysis, the impact of coupling could become significant if the comoving distance between primary and secondary lenses is sufficiently large. For instance, our investigation of MG\,J0414+0534 indicates a preference for the CDM model with LOS halos residing farther from the primary lens compared to WDM and MDM models. This circumstance may require a magnetic component in the deflection angle due to the coupling. Future observations of time delay could potentially break the degeneracy due to `extended multi-plane mass sheet transformation' (eMMST), stemming from the ambiguity in the redshifts of the LOS halos and the mass sheet and shear in their lens planes~\cite{inoue2023}.

Additionally, we omitted contributions from subhalos within the halo of the primary lens in our analysis. While we expect these contributions to be subdominant, their inclusion could enhance lensing shift powers, resulting in less stringent constraints on non-CDM models. Accounting for such contributions requires a thorough understanding of subhalo properties across various redshifts in non-CDM models. Cosmological simulations incorporating complex baryonic physics become indispensable for accurately estimating the perturbing effects of subhalos and LOS halos in non-CDM scenarios.

%%%%%%%%%%%%%%%%%%%%%%%%%%%
\section*{Acknowledgments}
%%%%%%%%%%%%%%%%%%%%%%%%%%%
This paper makes use of the following ALMA data: ADS/JAO.ALMA$\#$2013.1.01110.S., 2016.1.00281.S. ALMA is a partnership of ESO (representing its member states), NSF (USA) and NINS (Japan), together with NRC (Canada), MOST and ASIAA (Taiwan), and KASI (Republic of Korea), in cooperation with the Republic of Chile. The Joint ALMA Observatory is operated by ESO, AUI/NRAO, and NAOJ. This work is supported in part by the MEXT KAKENHI Grant Number 17H06359~(TS), JP21H05453~(TS), 23H04515~(TT) and the JSPS KAKENHI Grant Number JP19K03864~(TS), 19K03874~(TT), 23K17691~(TT).

%%%%%%%%%%%%%%%%%%%%%%%%%%%%%%%%%%%%%%%%%%%%%%%%%%%%%%%%%%%%%%%%%%%%%%%%%%%%%%%%%%%%%%%%%%%%%%%%%%%%%%%%%%%%%%%%%%%%%%%%%%%%%%%%%%%%%%%%%%%%%%%%%%%%%%%%%%%%%%%%%
\pagebreak
\appendix
\def\thesection{Appendix~\Alph{section}}
\section{Non-linear matter power spectrum \label{sec:app_nonlinear}}
\def\thesection{\Alph{section}}
%%%%%%%%%%%%%%%%%%%%%%%%%%%%%%%%%%%%%%%%%%%%%%%%%%%%%%%%%%%%%%%%%%%%%%%%%%%%%%%%%%%%%%%%%%%%%%%%%%%%%%%%%%%%%%%%%%%%%%%%%%%%%%%%%%%%%%%%%%%%%%%%%%%%%%%%%%%%%%%%%

In this paper, we employ the fitting formula for the non-linear matter power spectra in the WDM and MDM models based on the CDM one given by \cite{Takahashi:2012em} (and \cite{Inoue:2014jka}).
There are 8 parameters, $a_n$, $b_n$, $c_n$, $\al_n$, $\be_n$, $\ga_n$, $\mu_n$, $\nu_n$ in the formula. We use the same parameters as \cite{Takahashi:2012em} but adopt $a_n$, $c_n$ and $\ga_n$ from \cite{Inoue:2014jka}. 
The following is a summary of those parameters:   
\begin{align}
&\log_{10} a_n = 0.9221 + 2.0595 n_\mrm{eff} + 2.4447 n_\mrm{eff}^2 + 1.2625 n_\mrm{eff}^3 + 0.2874 n_\mrm{eff}^4 - 0.7601 C\,,\\[8pt]
&\log_{10} b_n = -0.5642 + 0.5864 n_\mrm{eff} + 0.5716 n_\mrm{eff}^2 - 1.5474 C\,,\\[8pt]
&\log_{10} c_n = 0.4747 + 2.1542 n_\mrm{eff} + 0.8582 n_\mrm{eff}^2 + 0.8329 C\,,\\[8pt]
&\al_n = \bigl| 6.0835 + 1.3373 n_\mrm{eff} - 0.1959 n_\mrm{eff}^2 - 5.5274 C \bigr|\,,\\[8pt]
&\be_n = 2.0379 - 0.7354 n_\mrm{eff} + 0.3157 n_\mrm{eff}^2 + 1.2490 n_\mrm{eff}^3 + 0.3980 n_\mrm{eff}^4 - 0.1682 C\,,\\[8pt]
&\ga_n = 0.2247 - 0.2287 n_\mrm{eff} + 0.9726 C - 0.0533 \ln\biggl(\frac{k}{h\mrm{Mpc}^{-1}} \biggr) \,,\\[8pt]
&\mu_n = 0\,,\\[8pt]
&\log_{10} \nu_n = 5.2105 + 3.6902 n_\mrm{eff}\,.
\end{align}
Here the effective spectral index $n_\mrm{eff}$ and the curvature $C$ are defined as
\begin{align}
n_\mrm{eff} + 3 = - \frac{d\ln \si^2(R)}{d\ln R}\biggl|_{\si=1}\,,\qquad C = - \frac{d^2\ln\si^2(R)}{d\ln R^2}\biggl|_{\si=1}\,,
\end{align}
or
\begin{align}
n_\mrm{eff} = I_1-3\,,\qquad C = 2I_1 + I_1^2 - I_2\,,
\end{align}
where
\begin{align}
I_1 = \frac{2}{k_\si^2} \int \frac{dk}{k}\, k^2 \De_\mrm{Lin}^2(k)\, e^{-(k/k_\si)^2}\,,\qquad I_2 = \frac{4}{k_\si^4} \int \frac{dk}{k}\,k^4 \De_\mrm{Lin}^2(k)\, e^{-(k/k_\si)^2}\,,
\end{align}
and the wave number $k_\si$ is determined so that $\si^2(k_\si^{-1}) = 1$.
Note that the variance $\si^2$ is computed by applying the Gaussian filter. The linear matter power spectrum $\De_\mrm{Lin}^2(k)$ is evaluated by the fitting formula given in \cite{Eisenstein:1997ik}.

We obtain the non-linear matter power spectrum $\De^2(k)\equiv k^3 P_\mrm{CDM}(k)/(2\pi^2)$ by substituting the above into the following expression \cite{Takahashi:2012em}:
\begin{align}
\De^2(k) = \De^2_Q(k) + \De^2_H(k)\,,
\end{align}
where
\begin{align}
&\De^2_Q(k) = \De_\mrm{Lin}^2(k)\, \Biggl[\frac{\bigl\{ 1+\De_\mrm{Lin}^2(k) \bigr\}^{\be_n}}{1+\al_n \De_\mrm{Lin}^2(k) } \Biggr]\, \exp\biggl(-\frac{(k/k_\si)}{4} - \frac{(k/k_\si)^2}{8}\biggr)\,, \\[8pt]
&\De^2_H(k) = \frac{1}{1 + \mu_n (k/k_\si)^{-1} + \nu_n (k/k_\si)^{-2}}\, \frac{a_n (k/k_\si)^{3f_1(\Om_m)}}{1 + b_n (k/k_\si)^{f_2(\Om_m}) + [c_n f_3(\Om_m)\,(k/k_\si)]^{3-\ga_n}} \,,
\end{align}
with
\begin{align}
f_1(\Om_m) = \Om_m^{-0.0307}(z)\,, \qquad f_2(\Om_m) = \Om_m^{-0.0585}(z)\,, \qquad f_3(\Om_m) = \Om_m^{0.0743}(z)\,.
\end{align}

\bibliography{WDM}

\end{document}